\documentclass[10pt]{amsart}

\usepackage{graphicx}
\usepackage{indentfirst,csquotes}
\usepackage{booktabs}

\usepackage{colortbl}
\usepackage[colorlinks=true,citecolor=red,linkcolor=blue,urlcolor=cyan,filecolor=magenta]{hyperref}

\usepackage{amssymb,amsthm,amsmath,amsfonts}

\newcommand{\Z}{\mathbb{Z}}

\title[Amino acid chiral amplification using MC]{Amino acid chiral amplification using Monte Carlo dynamic} 
\author[R. Cruz Simbron et al.]{Romulo Cruz-Simbrón$\,^{1,2,3}$}
\thanks{$\,^{1}$Department of Chemistry, University of  Colorado Boulder, Boulder, Colorado 80309, United States.\\
${}$\;\;\;\;\;$\,^{2}$Blue Marble Space Institute of Science, Seattle, Washington 98104, USA.\\
${}$\;\;\;\;\;$\,^{3}$Technology of Materials for Environmental Remediation Group (TecMARA), Faculty of Sciences, National  ${}$ \;\;\;\;\; University of Engineering, Av. Tupac Amaru 210, Lima-Peru.}
\address{Department of Chemistry, University of  Colorado Boulder, Boulder, Colorado 80309, United States.}
\email{romulo.cruz-simbron@colorado.edu}

\author[]{Gino Picasso$\,^{3}$}
\address{Technology of Materials for Environmental Remediation Group (TecMARA), Faculty of Sciences, National University of Engineering, Av. Tupac Amaru 210, Lima-Peru.}
\email{gpicasso@uni.edu.pe}

\author[]{Jose Cerda-Hern\'andez$\,^{4}$}
\thanks{$\,^{4}$Econometric Modelling and Data Science Research Group (EMDS), National University of Engineering, ${}$ \;\;\;\;\;  ${}$ \;\;\;\;\; Av. Tupac Amaru, 210, Lima-Peru.}
\address{National University of Engineering,  Av. Tupac Amaru, 210, Lima-Peru.}
\email{jcerdah@uni.edu.pe}

\begin{document}
\maketitle

\begin{abstract}
The present work focuses on the processes of chiral amplification that lead to the rapid growth of the enantiomeric excess in a solution, utilizing a lattice model and a suitable Glauber dynamics. The initial conditions stem from a racemic mixture or points near the racemic state. The aim is to understand the effect of some variables such as temperature, concentration and constants that define the interaction energies in the equilibrium concentration after the dynamic evolution of the system. Dynamic evolution involves a path towards phase equilibrium in a D-L-S system,  where D and L represent opposite chiral molecules and S represents their poorly soluble solvent. Our results, pertaining to the phase equilibrium of the D-L-S system employing amino acids, faithfully reproduce several experimentally observed outcomes documented in the literature.
Through simulations, we may understand how the system evolved over time, starting from a random configuration and moving toward an equilibrium state with the lowest possible potential energy. We were able to recreate phase diagrams that were strikingly close to those obtained experimentally by specifying an appropriate Glauber dynamics for the system. Finally, we will discuss some findings from the dynamics of the chiral amplification processes that were modeled.
\bigskip

\medskip

\noindent
\end{abstract} 

\section{Introduction}
The chiral asymmetry is ubiquitous in the biomolecules of the organisms, that is, there is a predominance of an enantiomer over its mirror image. Such enantiopreference is noticed in the chemical structures of sugar that form the nucleotides of DNA, or in the amino acids of the proteins. These facts have undoubtedly generated fundamental questions about how those chiral asymmetries arose spontaneously in the terrestrial biosphere, and in  what degree liquid solutions were involved in that scenario. It is also important to know if the exclusion of an enantiomer is a basic prerequisite for the origin, survival, propagation and evolution of living organisms, or if it is a secondary byproduct of the appearance of life \cite{lombardo2009thermodynamic}. From the perspective of the current development of chemical synthesis, the synthesis of enantiopure drugs is increasingly necessary and it is a challenge to synthesize an enantiopure compound at the lowest possible cost. With this approach, chiral amplification techniques play a major role in the current development of chemical synthesis.
\medskip

Several mechanisms have been proposed to explain this chiral asymmetry in physicochemical systems: a. Asymmetric autocatalysis: This scenario involves asymmetric autocatalytic chemical reactions; a phenomenon experimentally carried out by Kenso Soai \cite{soai2018asymmetric}. Asymmetric autocatalysis is the process by which a chiral reaction product is the catalyst of its own formation from achiral reactants. b. Attrition-enhanced deracemization of conglomerates: It refers to the process known as Viedma ripening \cite{engwerda2019attrition}, which consist in a deracemization process that involves a vigorous grinding of chiral crystals with racemization in solution, resulting eventually in the growing the only one of the two enantiomers through Ostwald ripening. c. Phase equilibrium: This third possibility focuses on the behavior of phase equilibrium between the enantiomorphs and a third component that acts as a relatively poor solvent \cite{klussmann2006thermodynamic}. d. In addition to the experimental studies, a series of computational calculations have been published addressing the origin of chirality. Klussman, experimentally, and Lombardo, theoretically, studied the enantiomeric excess of the solid and liquid phases in equilibrium, in relation to the global enantiomeric excess. The enantiomeric excess, $ee$, is defined in the Equation \ref{ee equation}.
\begin{equation}
	ee = \frac{|\alpha_D-\alpha_L|}{\alpha_D+\alpha_L}
	\label{ee equation}
\end{equation}
where $\alpha_D$ and $\alpha_L$ are the molar fractions of D and L respectively (e.i., if $\alpha_S$ is the molar fraction of the solvent, then $\alpha_D + \alpha_L + \alpha_S = 1$).  According to the Klussmann's results \cite{klussmann2006thermodynamic}, and in accordance with the rule of the phases, the composition of a proline solution in equilibrium with two solid phases (enantiopure and racemic) is fixed, and this composition is given by the composition at definite point. Only in cases where the enantiomeric excess is very low (where the excess of the enantiomer with the highest composition is too low to establish its own solid phase) or very high (where the concentration of the enantiomer with less abundance is insufficient to form a racemic compound) a variation of the composition of the solution with respect to the total enantiomeric excess could arise.

\section{Lattice model of the chiral amplification of amino acids}
In a seminal work, Lombardo, Stillinger, and Debenedetti \cite{lombardo2009thermodynamic} introduced a two-dimensional lattice model, known as Lombardo's model, to investigate the equilibrium of a ternary mixture comprising two enantiomeric forms of a chiral molecule (D and L) and a non-chiral liquid solvent (S). This model was developed to study the phase transition phenomenon, thermal equilibrium properties, and to calculate the behavior of the phases.
\medskip

The Lombardo's model start with a square lattice with $N$ rows and $N$ columns, and consequently  $N \times N$ sites. We denote the finite  square lattice  by  $\mathbb{L}_N^2=(\mathbb{V}_N,\mathbb{E}_N)$. Each enantiomer D and L can occupy one of the lattice sites of a $\mathbb{L}_N^2$ lattice with four available orientations. By adding the solvent (S) to the system, the Lombardo's model allows for 9 possible states for each molecule in the lattice (four kinds of spin for D and L, and one for S), each of which can occupy one site $i$ of the square lattice $\mathbb{L}_N^2$.  In the  mentioned  article a specific arrange of the lattice is called  {\it occupation status} or {\it spin configuration},  and it is denoted by  $\xi_i$ for each site $i \in \mathbb{V}_N$.  For each configuration $\xi=\{ \xi_i : i \in \mathbb{V}_N \}$ on the square lattice $\mathbb{L}_N^2$, Lombardo et al.  define the  potential energy of the   spin configuration  $\xi$ as follows 
\begin{equation}
	\varphi(\xi) = \nu_0N_0(\xi) + \nu_1N_1(\xi) + \nu_2N_2(\xi)
	\label{energie equation}
\end{equation}
where $N_0$ represents the number of nearest-neighbor pairs of enantiomorphic molecules, without taking into account their chiral type or their orientation, $N_1$ is the set of nearest-neighbor pairs of enantiomorphs of identical chirality and identical orientation, and  $N_2$ represents the set of neighboring square groups of four with alternating chirality and the same orientation (see \cite{lombardo2009thermodynamic} for more details). Notice that the sets $N_1$ and $N_2$ are disjoint while $N_1 \subset N_0$.   The Lombardo model cannot be solved exactly, primarily due to the geometrical constraints imposed by the finite-dimensional Euclidean space. However, it can be approximately solved in a relatively simple manner using the mean field approximation. This approximation treats a system of interacting particles as a system of non-interacting particles, wherein each particle only interacts with a ``mean field''  that captures the average behavior of the particles around it. This makes it a powerful method for exploring the behavior of complex many-particle systems that cannot be solved exactly.
\medskip

Obtaining closed formulas for the quantities  $N_0(\xi)$, $N_1(\xi)$ and $N_2(\xi)$ is a highly challenging combinatorics problem. In this regard, the numerical approximate solution studied by Lombardo assigns occupation probabilities to each point in the lattice, providing an approximation to the terms $N_0(\xi)$, $N_1(\xi)$ and $N_2(\xi)$. Lombardo also writes the entropy in terms of these occupation probabilities, which enables him to express the Helmholtz free energy in terms of the entropy and the potential energy.  Its next step involves the minimization of the free energy  with respect to the  parameters of the Lombardo's model. However, the minimization must  adhere to additional constraints: temperature equality of all the phases and that the chemical potential of each species must be the same in all the present phases. The chemical potential of species $i$ ($i =$ D, L, S) is defined as follow
\begin{equation}
	\mu_i = \frac{\partial F}{\partial n_i}
	\label{moles parcial}
\end{equation}
where $n_i$ is the number of molecules of type $i$. 

Numerical simulations of the Lombardo's  model  allows calculation of a ternary phase diagrams when obtaining the occupation probabilities in the points or regions of coexistence of the phases.  An outstanding characteristic of the phase diagram is the appearance of a pair of points characterized as triple points that imply the coexistence of a liquid phase enriched in one of the enantiomers with two solid phases: a racemic crystal and an enantiopure crystal. The theoretical results of Lombardo are qualitatively similar to the experimental results of Klussmann \cite{klussmann2006thermodynamic}. 
\medskip

Despite the promising results obtained by the mean-field approximation in Lombardo's model, effectively replicating Klussmann's findings, Lombardo suggests employing a more precise approach to better reproduce the equilibrium phase behavior of the ternary system and identify potential emerging properties not previously detected in the system. The well-established Monte Carlo simulation procedure offers a compelling alternative for this purpose.
\medskip

In this vein, the aim of this work is to introduce a new, straightforward model within Lombardo's framework, with the aim of reproducing experimental results through the implementation of a suitable Glauber dynamics. The subsequent section outlines the definition of our proposed model.

\section{The model}
\begin{figure}[!ht]
	\centering
	\includegraphics[scale=0.48]{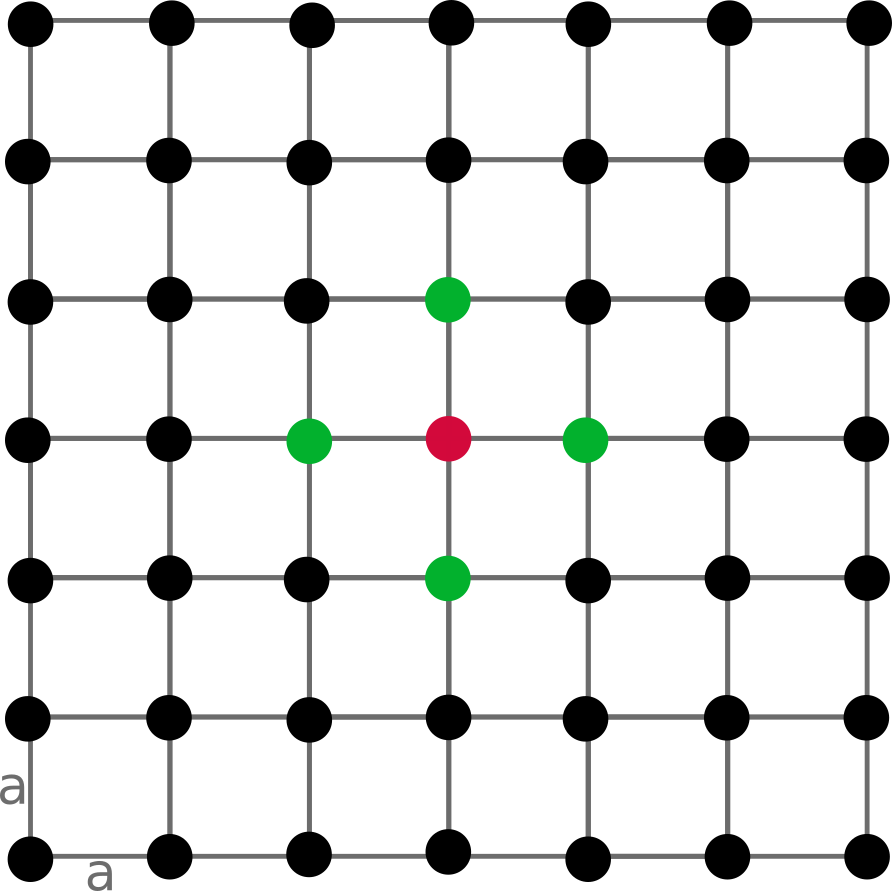}
	\caption{Lattice belonging to the set $\Z^2$ as an essential element of our model. Each point of the lattice corresponds to a certain molecule}
	\label{Z2_red}
\end{figure}

In this section, we define our model on  the classical two-dimensional lattices $\mathbb{L}_N^2$ (see Figure \ref{Z2_red}),   characterized  by the Gibbs measure  $\mu$ in terms of the Hamiltonian energy function. Each vertex  $i\in \mathbb{V}_N$  is conceived as being occupied by a chemical particle or random spin with specific properties. The chemical species or spins are restricted to three types: enantiomers L, enantiomers D, or solvent molecules S. Interactions between two chemical particles are restricted to their four nearest neighbors (see Figure \ref{Z2_red}). Considering that chemical particles are assumed to exist in three basic types, we take the set as the sample space
\begin{equation}
	\Omega = \{D,L,S\}^{\mathbb{V}_N}
	\label{configuration space}
\end{equation}

A configuration, denoted as $\sigma={ \sigma_i : i \in \mathbb{V}_N } \in \Omega$, represents a collection of spin values assigned to each vertex of the lattice. Thus, for example, a configuration in which there are no amino acid molecules and there are only solvent molecules will be represented as $ \sigma = \{ \sigma_i = S, \forall \; i\in \mathbb{V}_N \}$, where $\sigma_i$ is the spin associated with the  vertex $i$.  The term spin has its origin in the classical Ising square lattice model. This model represents a theoretical approach to understand the transition from ferromagnetic behavior to paramagnetic behavior. 
\medskip

\begin{figure}[!ht]
	\centering
	\includegraphics[scale=0.5]{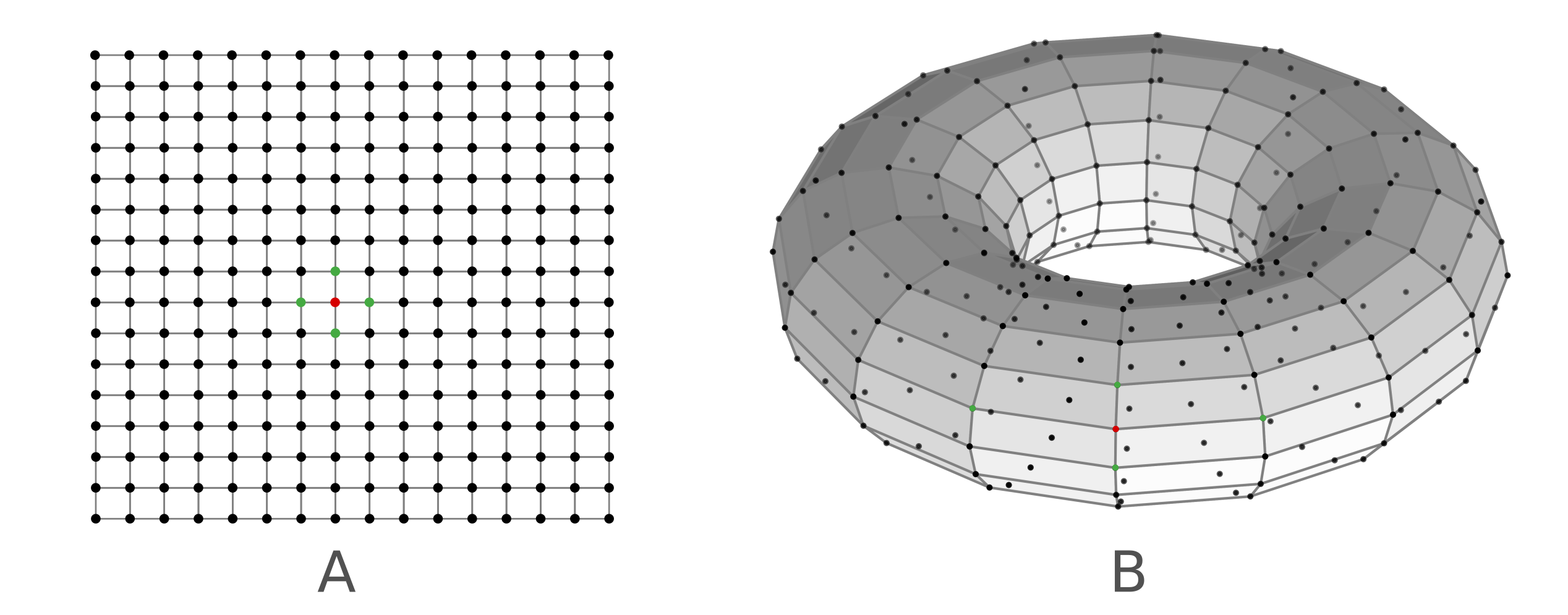}
	\caption{A: representation of a two-dimensional square lattice with side of 16 molecules. B: Torus $\mathbb{T}_{16}$ representing the periodic boundary conditions of the lattice in subfigure A.}
	\label{methodology_torus}
\end{figure}

In this work  the proposed model is defined on the square lattice $\mathbb{L}_N^2$.  Moreover, we have employed periodic boundary conditions, implying that a molecule situated on the boundary of the square lattice can interact with molecules positioned on the opposite boundary.  Mathematically, the square lattice  $\mathbb{L}_N^2$ with periodic boundary conditions can be visualized as a two-dimensional torus   $\mathbb{T}_N$, where  $N$ represents the size of the square lattice giving rise to the torus (Figure \ref{methodology_torus}). 

For any   $\sigma \in \Omega$, where  $\sigma_i \in \{ D, L, S \}$ is the spin value of configuration $\sigma$  at the site $i\in \mathbb{V}_N$, we define the following three functions that will allow us to define the energy of a given configuration $\sigma$,
\begin{itemize}
	\item $N_0(\sigma)$: total number of pairs of neighboring amino acids regardless of their chirality (LD, DL, DD, LL).
	\item $N_1(\sigma)$: total number of pairs of neighboring amino acids with equal chirality (DD, LL).
	\item $N_2(\sigma)$: total number of amino acid quartets that have alternating chirality forming a square,
	$$
	\begin{array}{cc}
		L &D \\
		D & L
	\end{array}
	\;\;\; \mbox{or} \;\;\;
	\begin{array}{cc}
		D &L \\
		L & D
	\end{array} 
	$$
\end{itemize}
Using these functions,  the formal  potential energy (Hamiltonian) of a given configuration $\sigma$ of the chemical species, $\varphi(\sigma)$, will be defined according to the following equation
\begin{equation}
	\centering
	\varphi(\sigma) = \nu_0N_0(\sigma) + \nu_1N_1(\sigma) + \nu_2N_2(\sigma)
	\label{energie equation in paper}
\end{equation}
where the coefficients $\nu_0$,  $\nu_1$ and $\nu_2$ determine the importance of each type of interaction in the final potential energy. Here, the quantities $\nu_0, \nu_1$ and $\nu_2$ are negative.  For a physical system like a liquid-solid equilibrium like the one we are studying, the Boltzmann distribution on $\Omega$ (also known as the Gibbs distribution) is the appropriate stationary distribution, see Ref. \cite{friedli2017statistical}. The Gibbs distribution $\mu$ on $\Omega$   is given by 
\begin{equation}
	\mu(\sigma) = \dfrac{1}{Z}e^{-\beta\varphi(\sigma)}, \;\; \sigma\in \Omega
	\label{Gibb_distribucion}
\end{equation}
where $\beta=\frac{1}{\kappa_B T}$,  $k_B$ is Boltzmann's constant and $T$ is the temperature. The partition function $Z$ is the normalization constant required for $\mu$ to represent a probability distribution and it  is given by
\begin{equation}
	Z = \sum_{\sigma \in \Omega}e^{-\beta\varphi(\sigma)}
\end{equation}

\section*{Non-classical nucleation process}
In this work, unconventional   crystallization processes are also studied using our model, and involve the theory of prenucleation clusters (PNC) proposed by the Helmut Cölfen group(see  Ref. \cite{gebauer2011prenucleation}). PNCs are part of an unconventional and opposite crystallization processes to the classical nucleation  theory (CNT) (see  Ref. \cite{kellermeier2012amino}). The PNCs are presented under a reversible equilibrium with their components dissolved in the solution, which is characteristic of their association constants and is directed to a minimum of Gibbs energy given by \ref{energie equation in paper}. In addition, PNCs are stable in low saturation regimes of solutions, where there is no thermodynamic force for crystallization according to the classical theory of nucleation. By increasing the concentration the system is directed to an aggregation of the clusters and to a precipitation. The experimental study of NCPs is a challenge, especially since the clusters are small and even present only in very small concentrations. Techniques such as the analytical ultracentrifugation (AUC), cryo trasmission electron microscopy (cryo TEM) or electrospray ionisation mass spectrometry (ESI MS) have shown good results in the study of the PNC; however, the requirements of speed and robustness have not yet been met for the characterization of PNC.
\medskip

Kellermeier \cite{kellermeier2012amino} shows that amino acids also form clusters that vary from two to eleven units of monomer. He finds that tetratemeros or trimers occur in greater abundance and that as the concentration increases, there is a limit to the average cluster size. Kellermeier studies the range of concentrations from 0.0001 M to 1 M and also in the supersaturation regime with 1.56 and 1.82M (solubility of arginine is 1.3M). He finds that in the case of solutions below the limit of solubility there is an average limit cluster size. It also finds that the cluster distribution does not depend on whether it is type L or type D. There is also no change when a racemic mixture or enantiopure mixture is present, at least in the average cluster size or the highest abundance. However, Nemes et al. \cite{nemes2005amino} shows that other amino acids have a difference in their MS spectra when they have a racemic or enantiopure mixture. 
\medskip

The lattices models and the Monte Carlo dynamics are useful computational tools that are increasingly used in the fields of chemical process simulation (see for example Refs. \cite{lombardo2009thermodynamic}, \cite{hatch2010chiral}).  In front of the processes of molecular dynamics and quantum calculations, its advantage lies in being able to establish a global relationship in interactions  being able to approach a multitude of stages of the processes much higher at a very low computational cost (see Ref. \cite{marro2005nonequilibrium}). Although the details of the interactions are ignored in a lattice model,  the collective properties or relevant macroscopic quantities of the system  that can arise are largely correlated with the experimental results,  since when there is a break in the symmetric, as are the processes of chiral amplification, systems as a whole present properties beyond the individual properties of molecules (see Refs. \cite{sole1999phase},  \cite{sole2011phase}).  The numerical  results of  our model presented here, establishes a model also of non-conventional crystallization process.
\medskip

The lattice model in conjunction with a suitable Glauber dynamics, described in the following section, manages to quantitatively reproduce the results of Klussman regarding the chiral amplification phenomenon that it presents in its experimental results. This prediction allows to establish the preponderant role played by certain variables such as temperature, concentration or intermolecular interaction constants that determine the overall evolution of the system.  The crystallization process of the amino acids is focused from the point of view of the generation of prenucleation clusters and this approach is corroborated by the theoretical results that support the experimental results, especially with regard to the magic numbers.

\section{Glauber dynamics}
Below, we define the Glauber dynamics for the our model over the configuration space $\Omega$. These dynamics are governed by a reversible Markov chain with stationary distribution $\mu$ defined in Eqn. \ref{Gibb_distribucion}.
\medskip

To define the Glauber  dynamic on our system we use the \textbf{Metropolis–Hastings} (M-H) algorithm which is part of a much larger set of algorithms called \textbf{Markov chain Monte Carlo methods}. The M-H algorithm is a widely used procedure for sampling from a specified distribution on a large finite set. In our case,  starting from a random configuration, and through a sequence of permutations of nearest-neighbor amino acids, we  will use  this algorithm  to obtain a sampling of the Gibbs distribution $\mu$,  defined  in  Eqn. \ref{Gibb_distribucion}, and compute some relevant macroscopic quantity of the system. Thus,  the  M-H algorithm  give us  the transition probability $P_{\sigma \rightarrow \sigma'}$ to transitioning from an initial configuration $\sigma$ to another $\sigma'$ in a single time unit, in order to obtain  a  sample  according to  the Gibbs distribution $\mu$    (see Ref. \cite {kumar2020nonequilibrium}). The transition probability $P_{\sigma \rightarrow \sigma'}$ is provided by the following expression:
\begin{equation}
	P_{\sigma \rightarrow \sigma'} = \left\lbrace
	\begin{array}{ll}
		\frac{\mu(\sigma')}{\mu(\sigma)}=  e^{-\beta\Delta\varphi} &  \; \Delta\varphi\geq 0  \\
		1 & \; \Delta\varphi < 0  ; 
	\end{array}
	\right.
	\label{Conditional Metropolis-Hasting}
\end{equation}
where  $\Delta\varphi = \varphi(\sigma')- \varphi(\sigma)$.  The formula  (\ref{Conditional Metropolis-Hasting}) has a simple interpretation: from $\sigma \in \Omega$,  choose $\sigma' \in \Omega$ with probability  $P_{\sigma \rightarrow \sigma'}$;  if  $\Delta\varphi < 0$  move to  $\sigma'$; if  $\Delta\phi\geq 0$, flip a coin with  success probability equal to $e^{-\beta\Delta\varphi}$  and move to $\sigma'$ if success occurs;  in other cases, stay at  $\sigma$. Notice that it is impractical to compute the partition function, which is a sum over all  $3^{|V|}$ configurations. Fortunately the M-H algorithm allows us to draw random samples from a Gibbs distribution $\mu$ without knowing the normalization factor $Z$. Therefore, the steps of the M-H algorithm applied to our model are as follows:

\begin{figure}[!t]
	\centering
	\includegraphics[scale=0.3]{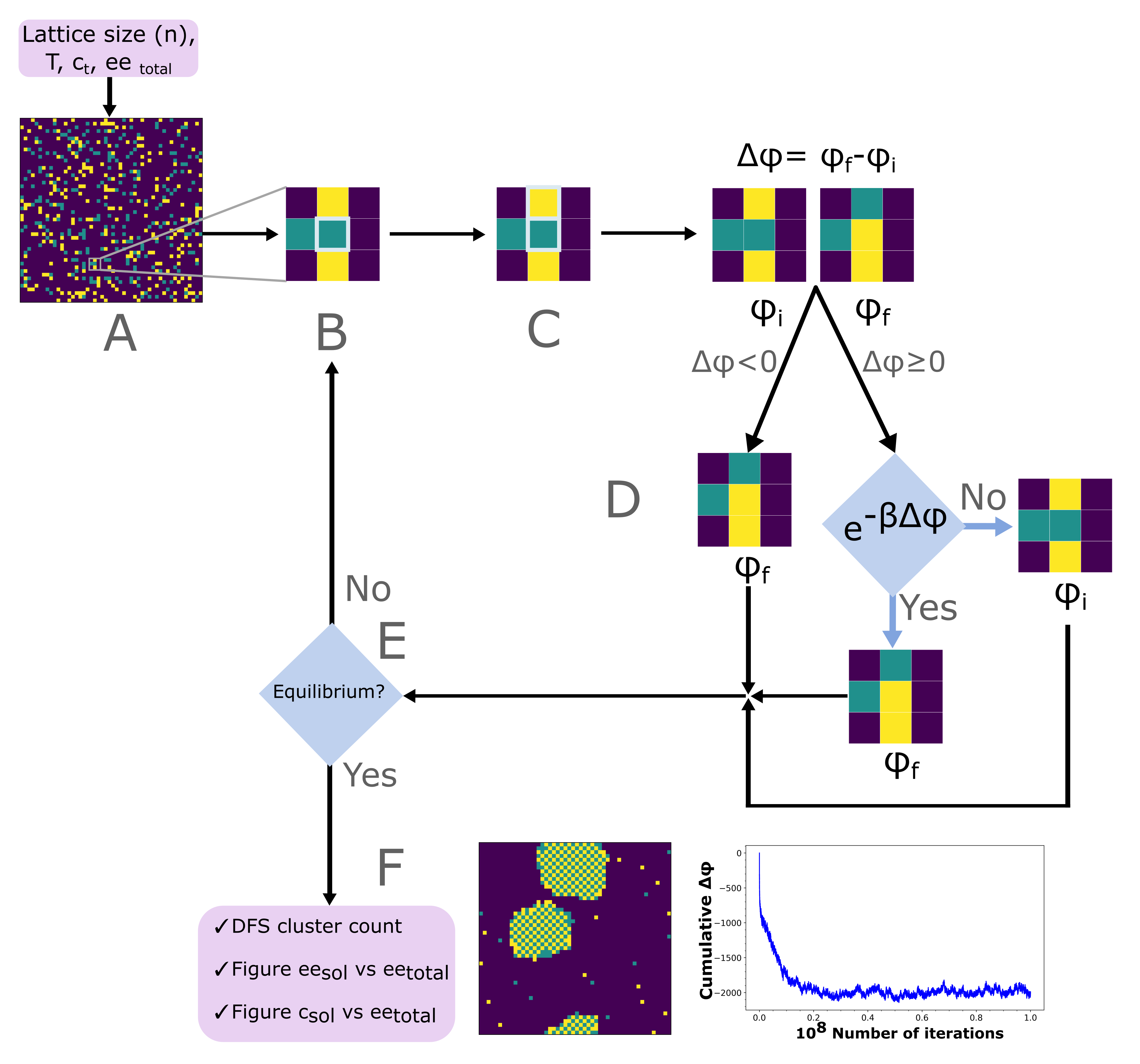}
	\caption{M-H algorithm applied to the crystallization process of chiral amino acids. A: Random lattice of amino acids L or D in a solvent S, B: Random selection of one element of the lattice. C: Selection of one of the four neighbors. D. Conditionals that define the exchange of positions. E: Conditional that ensures reaching equilibrium in the evolution of the system. F: Evaluation of the results.}
	\label{algorithm}
\end{figure}

\begin{enumerate}
	\item Generate a square lattice with molecules D and L in random positions. To accomplish this, establish the size of the lattice, denoted by $n$, as well as the predetermined number of molecules D and L within the model. These parameters will determine the system's total enantiomeric excess as well as its overall concentration.
	
	\item Select a vertex $i$ from  $\mathbb{V}_N$  at random,  ensuring that each element has an equal chance of being chosen (uniform random selection). At the chosen vertex, either an amino acid or the solvent may be present. \label{return_mole_random}
	
	\item Select a nearest neighbor uniformly using uniform random selection. Each molecule has four neighboring molecules.
	
	\item Calculate the potential energy change, $\Delta\varphi$, that would cause the permutation of positions of the initially selected molecule and its neighbor.
	
	\item  If  $\Delta\varphi $ is positive, the algorithm will check if the value of  $e^{-\beta \Delta\varphi}$   is greater than a random uniform number generated in the interval  $(0,1)$. If this condition is satisfied, the positions of the molecules will be swapped.  Otherwise, no changes will be made. \label{criterion_change}
	
	\item Go back to step \ref{return_mole_random} and repeat the process until the maximum number of iterations is reached.
\end{enumerate}

In Figure \ref{algorithm}, we show the flow chart of our algorithm for the Glauber dynamics on the square lattice system $\mathbb{L}_N^2$. It is important to highlight that the energy change from one configuration to another in a single time unit of the algorithm depends only on the immediate surroundings of the considered pair. To ensure that equilibrium is reached, the number of iterations (``\textit{mixing time}'') must be large enough. This allows the Glauber dynamics to visit all configurations of the sample space with high probability, ensuring convergence towards a state in which the energy does not vary beyond its natural variability, which is associated with the heat capacity of the system. (see for details  Ref. \cite{wang2001efficient}). In our lattice model with  $n=50$ rows and $n=50$ columns, a number of iterations of $10^8$ ensures that we reach the equilibrium in all trials performed. After completing the $10^8$ iterations, the system evolves again for $10^8$ more iterations, taking a sample every 100,000 iterations. This results in 1000 samples for each simulation. From each of these samples, the key variables for the system, which we define below, are calculated.  Consequently,  an average state of the system is obtained from each simulation. To account for the variability associated with the initial conditions, each simulation is repeated 10 times with a different random lattice and the results of the variables are averaged. This also allows us to obtain the standard deviations of each variable.
\medskip

The following variables are defined for this study:
\begin{equation}
	x^D_s = \dfrac{D_s}{S},\;
	x^L_s = \dfrac{L_s}{S},\;
	x^D_t = \dfrac{D_t}{S},\;
	x^L_t = \frac{L_t}{S}
	\label{Eq-fracciones}
\end{equation}

\begin{equation}
	ee_{sol} = \frac{x^D_s-x^L_s}{x^D_s+x^L_s} \; 100
	\label{Eq-exceso_solucion}
\end{equation}

\begin{equation}
	ee_{total} = \frac{x^D_t-x^L_t}{x^D_t+x^L_t} \;100
	\label{Eq-exceso_total}
\end{equation}

\begin{equation}
	c_s = \frac{D_s+L_s}{S},\;c_t = \frac{D_t+L_t}{S}
	\label{Eq-concentracion}
\end{equation}
where $D_s$ and $L_s$ represent the number of D and L molecules, respectively. $S$ represents the number of solvent molecules. The variables $x^D_s$ and $x^L_s$ refer to the fraction of D and L molecules in solution compared to the solvent molecules. $ee_{sol}$ and $ee_{total}$ represent the enantiomeric excess in solution and total, respectively. 

\section{Programming of the algorithms used}
The M-H algorithm (\ref{Conditional Metropolis-Hasting}) has been implemented in the C++ language. This language allows us to execute the simulations in an average time of approximately 2 hours per simulation and a RAM memory consumption of between 10 to 15 GB during its execution. \\
\medskip

To obtain the aforementioned variables, it is very important to have an algorithm that allows us to efficiently count and characterize the clusters of molecules formed. In the present work we have implemented the Depth First Search algorithm (DFS) in the programming language C++ \cite{imai1986efficient}. This algorithm, as well as the C++ implementation of the M-H algorithm and sub algorithms used during this work, can be found on the GitHub platform \href{https://github.com/romulo-cruzsimbron/quirality_montecarlo}{Code repository}.

\section{Results}
In this section we will detail the results of the computational simulations carried out. For these simulations, the lattice size has been defined as 50 x 50, that is, 2500 molecules distributed in a square lattice belonging to $\mathbb{L}_{50}^2$ with 50 molecules per edge. Our simulations model the effect of the concentration of amino acids, the total enantiomeric excess, the temperature and the value of the constant $\nu_1$  on  the enantiomeric excess in solution, the concentration in solution, the distribution of clusters in solution, the chiral amplification processes and the formation of aggregates of defined sizes and geometries.

\subsection{Effect of amino acid concentration}
These simulations seek to determine how our square lattice model reproduces the solubility property of a supersaturated racemic system, as well as the nucleation mechanism. For this purpose, the total concentration of amino acids, $c_t$, has been gradually increased and the concentration of amino acids remaining in solution, $c_s$, is determined. The temperature is kept fixed during the evolution of the system ($T = \frac{1}{\beta}$ = 0.8 to 1.9) and the simulations are stopped when the system no longer shows a decrease in potential energy, that is, when the sum of all the $\Delta\varphi$ has reached a constant value. For these simulations, the values of the constants $\nu_0 = -1$, $\nu_1 = -2$ and $\nu_2 = -5$ have been used, values taken from the work of Lombardo et al. \cite{lombardo2009thermodynamic}. As the systems are racemic, $ee_{total} = 0$ for all the simulations in this section.

\begin{figure}
	\centering
	\includegraphics[scale=0.45]{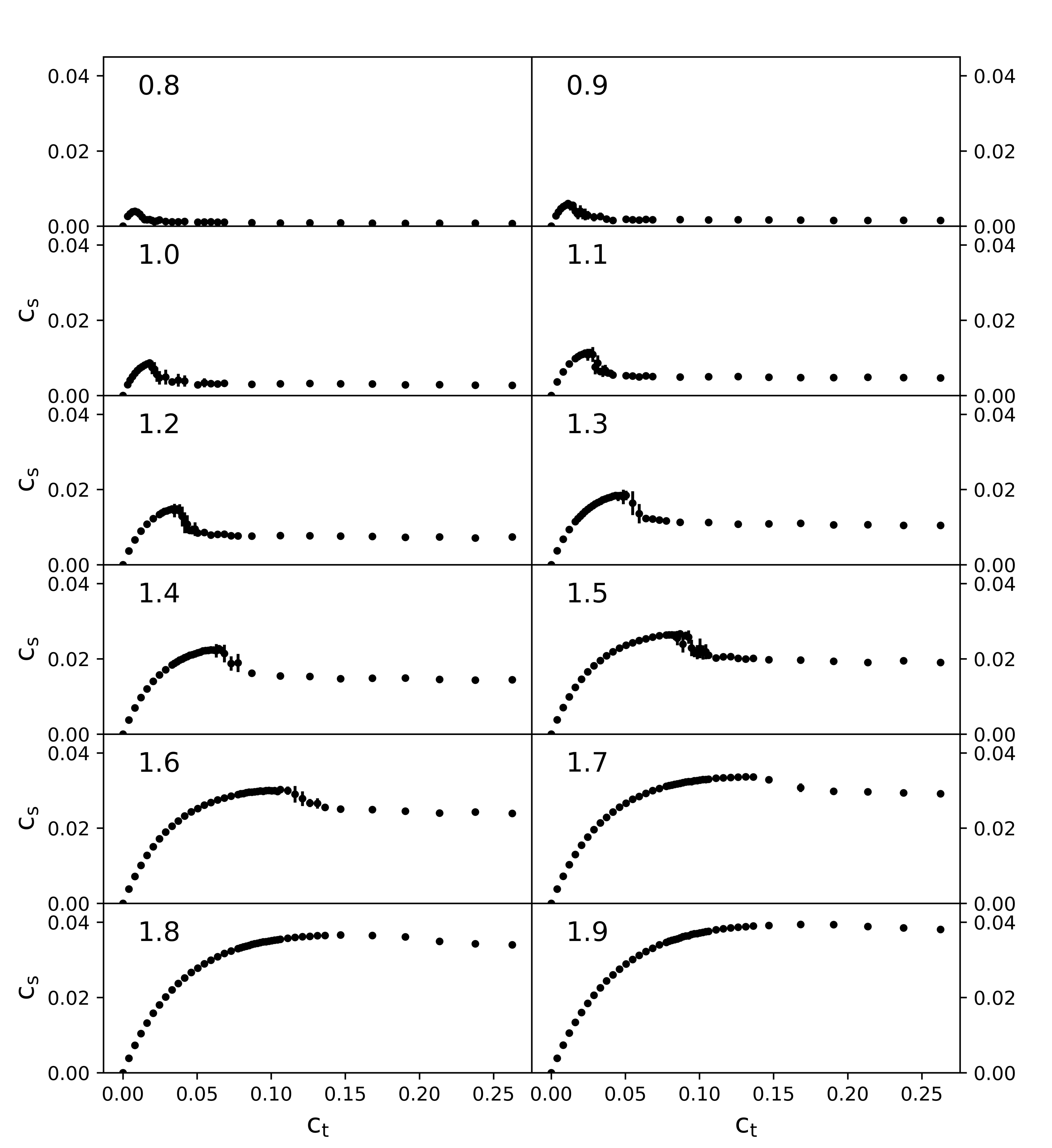}
	\caption{Effect of total amino acid concentration, $c_t$, on concentration in solution, $c_s$. The reduced temperature $T$ has been indicated within each subfigure.}
	\label{Effe_ConcentracionTotal}
\end{figure}

Figure \ref{Effe_ConcentracionTotal} shows that as the total concentration of amino acids increases, the concentration in solution goes through a maximum and then decreases to an almost constant value. The error bars are shown on each curve, result of having performed ten repetitions as indicated in the methods. The small variability in the results indicates that we are effectively facing a minimum of energy in the system that evolves by virtue of the Monte Carlo dynamics, the chosen constants $C_i$ and the temperature.  The region of nearly constant solution concentration corresponds to the saturation region. Our model thus evolves from an unsaturated state (subsaturation region), in which the saturation concentration value has not been exceeded, towards a supersaturated state. The system then reaches a state of constant saturation. 

\begin{figure}
	\centering
	\includegraphics[scale=0.45]{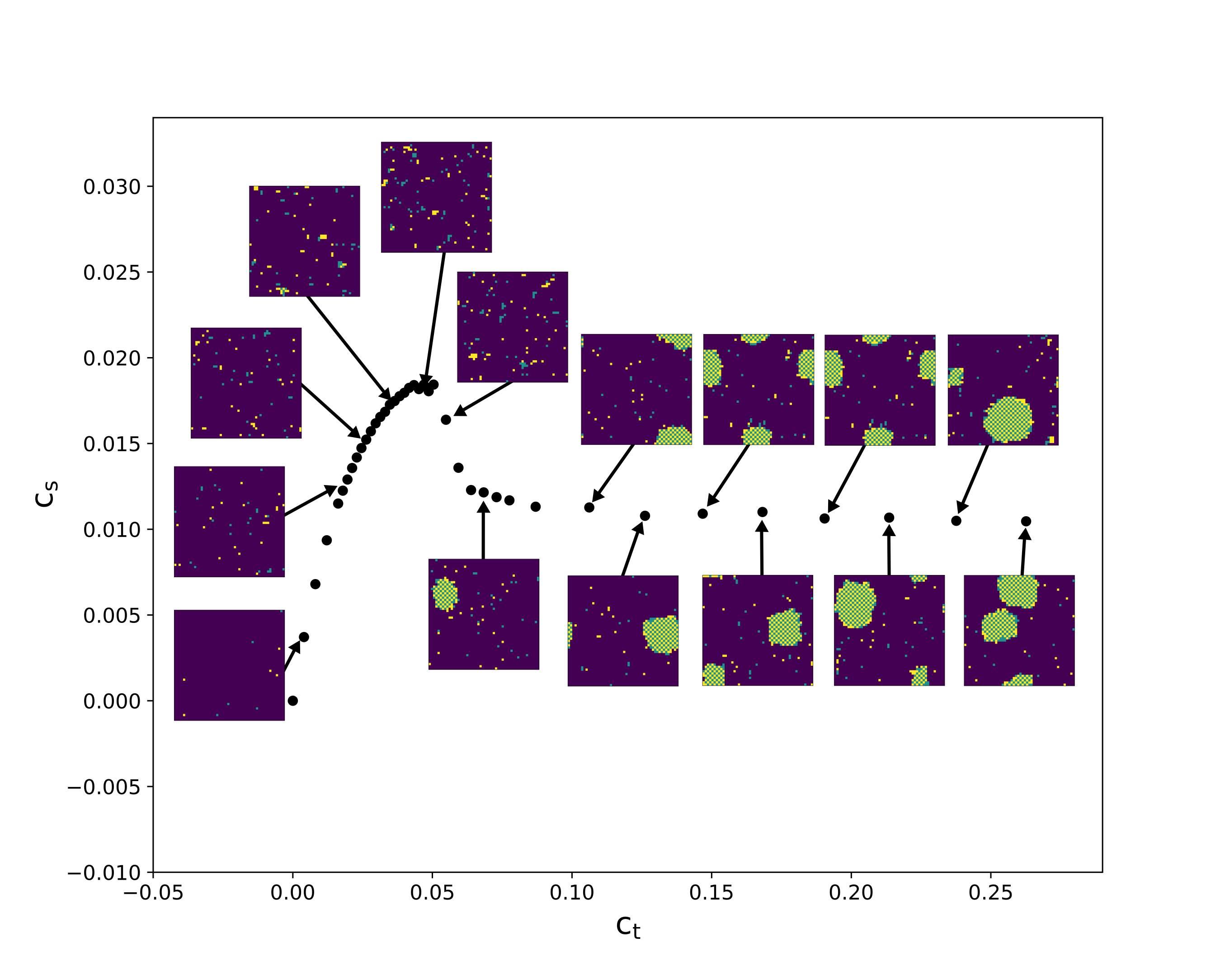}
	\caption{Square lattice showing the crystallization of amino acids in the system at $T$ = 1.3 as the total concentration of amino acids increases.}
	\label{Effe_ConcTotal_T13_Cristales}
\end{figure}

Taking the example of the system's evolution at a temperature of $T = 1.3$, as shown in Figure \ref{Effe_ConcTotal_T13_Cristales}, we observe that increasing the total concentration of amino acids leads to the formation of small aggregates of varying sizes, which coexist until the system go beyond the peak. Then, clusters of amino acids on the order of 100-150 in size are formed. Subsequently, the clusters grow, but the concentration in solution remains relatively constant. These clusters have been observed experimentally, with certain amino acids exhibiting clusters of a specific size with exceptional abundance \cite{scutelnic2018structure}, which is these sizes have been called ``magic numbers'' \cite{concina2006formation}. Mass spectrometry has observed glycine dimers \cite{friant2010glycine,zimbitas2019investigation}, alanine dimers and tetramers \cite{malar2018structural} and protonated serine dimers or octamers \cite{counterman2001magic,nanita2006serine,jordan2020effects}. The stability of the glycine dimer has been attributed to the free energy of formation of the pair $G_2$ (Equation \ref{glicina dimer}). Jordan \cite{jordan2020effects} determined that the serine clusters observed during their electrospray ionization mass spectrometry analysis actually exist in solution and are not the result of the spray evaporation process during the analysis. One of the general characteristics shared by the various models of the structure of serine octamers is the ability of these clusters to form a succession of hydrogen bonds that manages to stabilize them with respect to other clusters of different sizes \cite{scutelnic2018structure, jordan2020effects,counterman2001magic,schalley2002unusually}. A surprising aspect of serine octamers is that in a racemic solution they generate clusters with a strong homochiral preference, even at concentrations as low as 100 $\mu M$. In the Supplementary Material section we can see the other lattice for the other temperatures ( Figures S1 to S12).

\begin{equation}
	\begin{split}
		&G_{(ac)} + G_{(ac)} \rightarrow G_{2(ac)} \\ 
		&\Delta E = -15.2 \;kcal.mol^{-1} \\ 
		&\Delta G = -4.0 \;kcal.mol^{-1}
	\end{split}
	\label{glicina dimer}
\end{equation}

\begin{figure}[!ht]
	\centering
	\includegraphics[scale=0.5]{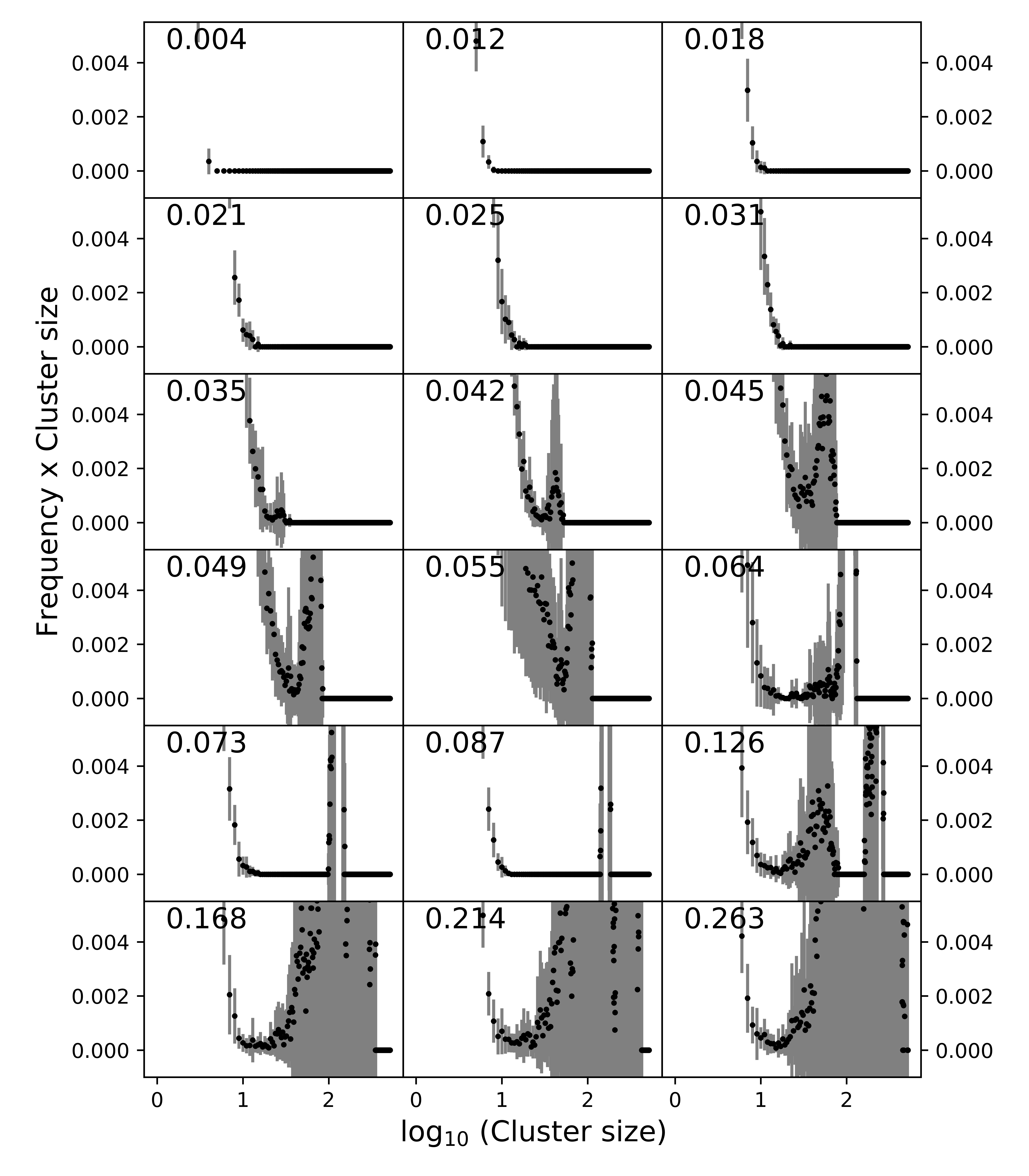}
	\caption{Size distribution in the simulations performed at $T$ = 1.3. Within each graph the value of the concentration in solution $c_t$ has been indicated}
	\label{DistribucionTotal_T1.3}
\end{figure}

Figure \ref{DistribucionTotal_T1.3} shows the evolution of the distribution of the size of the clusters as the concentration in solution increases. This Figure refers to the system at T = 1.3. The abscissa axis has been transformed to a logarithmic scale to appreciate the changes in the smaller clusters. The frequency of appearance of the clusters is defined as:
\begin{equation}
	\centering
	Frequency(n) = \frac{ \textnormal{Number of particles with size i}}{\sum_{i = 1} \textnormal{Number of particles with size i}}
\end{equation}
Where the sum of the denominator goes over all the sizes present in the system. The value of the Frequency multiplied by the size of the cluster has been taken as the y-axis of our graph , in order to correctly appreciate the evolution of the largest clusters. The values of $c_t$ have been placed in each subgraph to indicate the total concentration of the system in each specific subgraph.
\medskip

Figure \ref{DistribucionTotal_T1.3} shows that as we increase the total concentration of amino acids, the frequency of larger clusters increases. The data present an intrinsic variability that has its maximum in the concentration $c_t=0.055$ as it can be observed by the large error values in cluster size below 10 (1 in log scale). The main question is at what concentration value $c_t$ does the nucleation begin? . We notice that there are clusters in the concentration of $c_t=0.021$. If we look at Figure \ref{Effe_ConcTotal_T13_Cristales}, in this concentration is near to the supersaturation region. After this point, and as the concentration in solution increases, larger crystals are formed (growing), even though their variability is relatively large. In $c_t= 0.064-0.168$, the crystals formed are very stable and no longer present an appreciable variability. In addition to this, the size distribution is unimodal, that is, they present a single maximum size. When the concentration increases above $c_t=0.168$, the variability in the size of the crystals increases and a very wide distribution of sizes is presented that becomes more extensive as the concentration increases. What would be happening is that in $c_t=0.168$ the concentration is so high that it is possible to generate a greater number of crystallization nuclei. For example, in Figure \ref{Effe_ConcTotal_T13_Cristales} we notice that from $c_t=0.20$ two crystals are already shown (remember that the lattice is periodic).
\medskip

As Gebauer \cite{gebauer2008stable} has pointed out, the better stability of the clusters against the monomers originates a nucleation process very different from the one predicted by the classical nucleation theory. This type of behavior in a crystallization process has also been observed during the crystallization of calcium carbonate \cite{picker2012multiple} whose nucleation process has been postulated by Gebauer to correspond to a process of formation of prenucleation clusters which has its origin in the stability of the dimer CaCO$_3$-CaCO$_3$. Figure \ref{DistribucionTotal_T1.3} indicates that at low concentrations of amino acids between 0.31 and 0.35 there is a preference for cluster sizes between 10 and 30 nm. In this range of concentrations the system is in the region of supersaturation \ref{Effe_ConcTotal_T13_Cristales} but we have not yet passed the maximum concentration. 

\subsection{Solubility of a suppersaturated racemic mixture}
At this point the curiosity led us to know what is the behavior in our model of the saturation concentration (or solubility) with respect to temperature. This question will allow us to determine if the solubility in our models fits the van't Hoff equation or the Hildebrand equation \cite{grant1984non}. To make the solubility curves we have established the average of the last four points as the saturation concentration of each graph in Figure \ref{Effe_ConcentracionTotal}. These points represent the region of constant saturation. From these values we can graph the van't Hoff plot (Equation \ref{VanHoff}) or the Hildebrand plot (Equation \ref{Hildebrand}) following the equationss:
\begin{equation}
	ln(c_s) = A\;\left(\frac{1}{T}\right) + B 
	\label{VanHoff}
\end{equation}
\begin{equation}
	ln(c_s) = C\;ln(T) + D
	\label{Hildebrand}
\end{equation}

\begin{figure}
	\centering
	\includegraphics[scale=0.3]{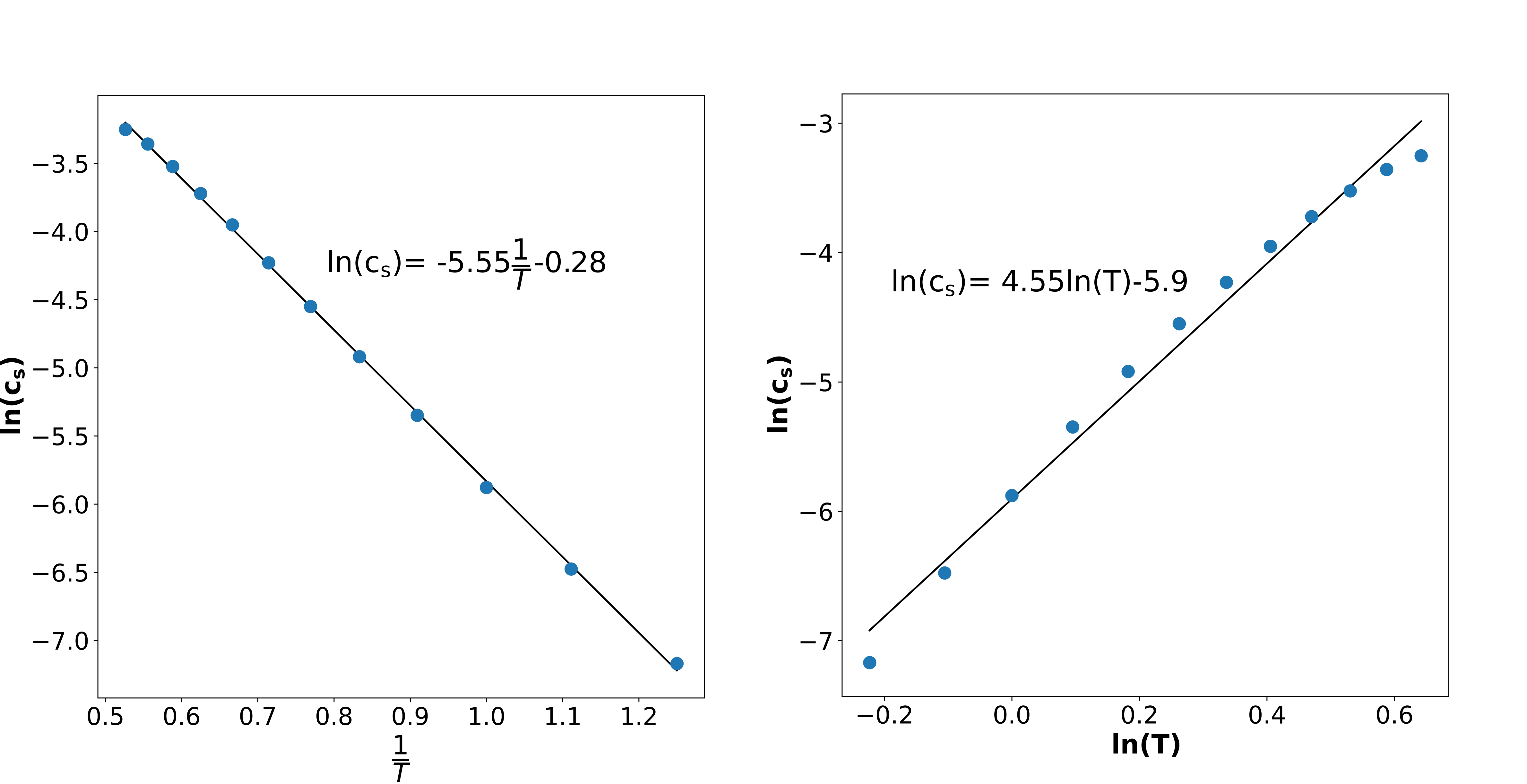}
	\caption{van't Hoff and Hildebrand solubility plots.}
	\label{SolubilidadC1-1}
\end{figure}

Where A, B, C and D are constants. Figure \ref{SolubilidadC1-1} shows that the data linearly fit a van't Hoff equation better than a Hildebrand equation. Even though the fit is qualitatively very good towards the van't Hoff equation, it is important to note that the curve seems to better follow a sigmoidal curve since that, at small values of $\frac{1}{T}$, the concentration in solution is above the linear trend line and at high values of $\frac{1}{T}$ the concentration $c_s$ is below the linear trend line.

\subsection{Homochiral preference in clusters in a supersaturated racemic mixture}
\begin{figure}
	\centering
	\includegraphics[scale=0.6]{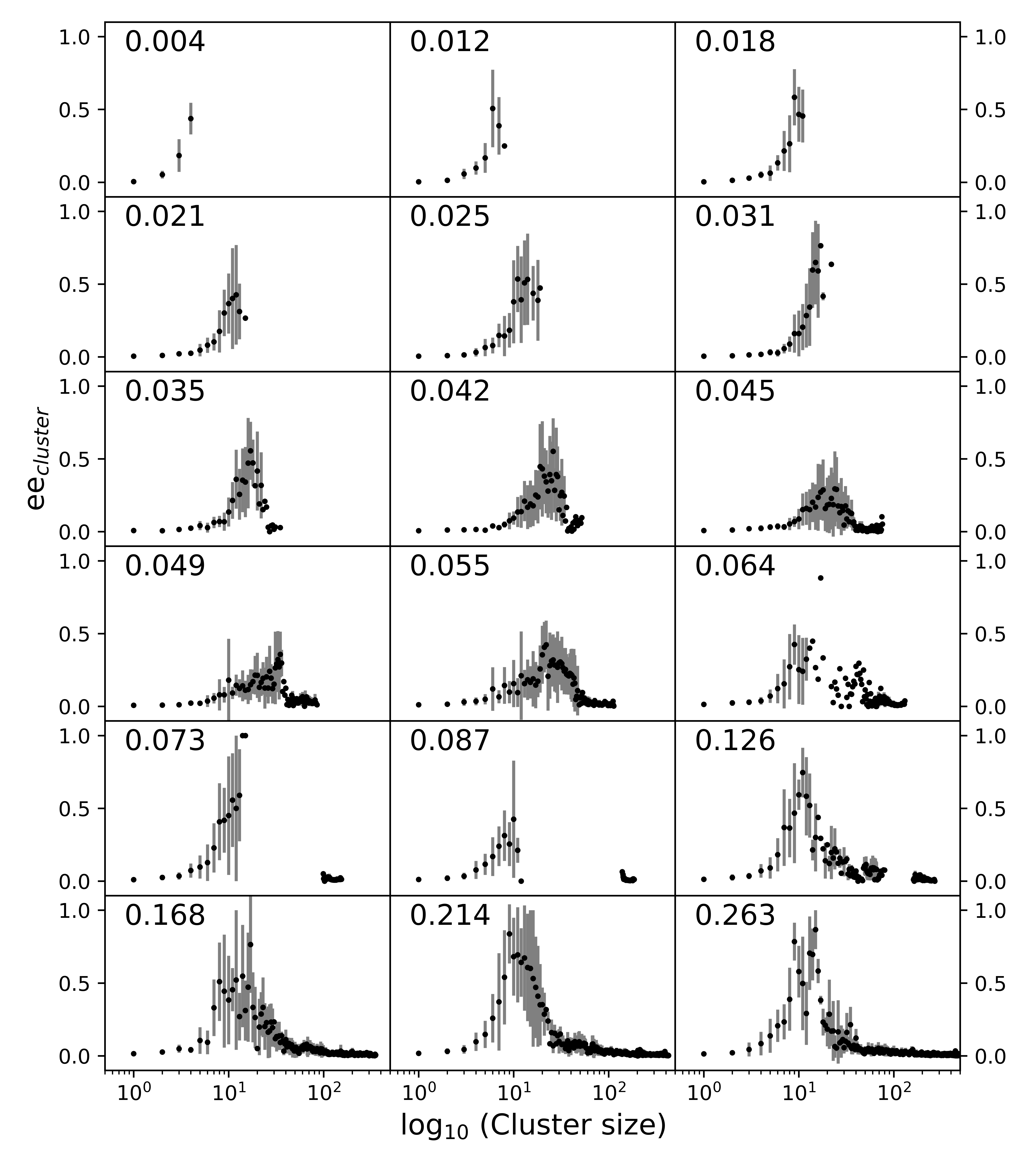}
	\caption{Chirality of the clusters formed by increasing the total amino acid concentration.}
	\label{Quiralidad_Clusteres}
\end{figure}

As we have briefly commented in one of the previous paragraphs, protonated serine octamers surprisingly show a homochirality preference in a racemic solution. To investigate this homochiral preference in the clusters of our system at $T = 1.3$ we have written a C++ program to obtain the average homochirality of each cluster size. For example, clusters of size 2 were counted and the enantiomeric excess of that cluster, $ee_{cluster}$, was determined. As we have 10 repetitions for each concentration condition, the $ee_{cluster}$ have been averaged considering their absolute value. Taking the absolute values is done because the homochiral preference can be towards both the L isomer and the D isomer. Figure \ref{Quiralidad_Clusteres} shows the results of this analysis. We note that at all concentrations, the dimmers (cluster size = 2) are on average racemic ($ee_{cluster} = 0$), but as the cluster size increases the model shows a homochiral preference of the clusters. This may be evident because the $\varphi$ function favors homochiral pairs energetically, however, something that is very interesting is that this homochiral preference has a maximum whose position depends on the total concentration of amino acids in solution. Thus, for low concentrations such as $c_s = 0.004-0.031$, the clusters can have an enantiomeric excess between 0.5 and 1 in sizes from 6 to 15 monomers.
\medskip

\subsection{Effect of $ee_{total}$ on enantiomeric excess in solution  $ee_{sol}$}
As we have seen in the previous section, in a racemic system the concentration in solution reaches an almost constant value when the concentration is high enough. We now wonder about the effect of the total enantiomeric excess at this high concentration can affect the enantiomeric excess in solution. This question is closely related to the main objective of the work, if our model manages to reproduce the chiral amplification that Klussmann \cite{klussmann2006thermodynamic} has studied experimentally for amino acids. The studied system maintains the parameters at $\nu_0 = -1$, $\nu_1 = -2$ and $\nu_2 = -5$ with a total concentration of amino acids, $c_{total}$, of 0.25. We chose this concentration as it allows to be in the saturation region in each simulation. The total enantiomeric excess, $ee_{total}$, has varied between 0 and 1 and the cases of temperature at $T =1.1,1.3,1.5$ and $1.7$ have been studied.

\begin{figure}
	\centering
	\includegraphics[scale=0.5]{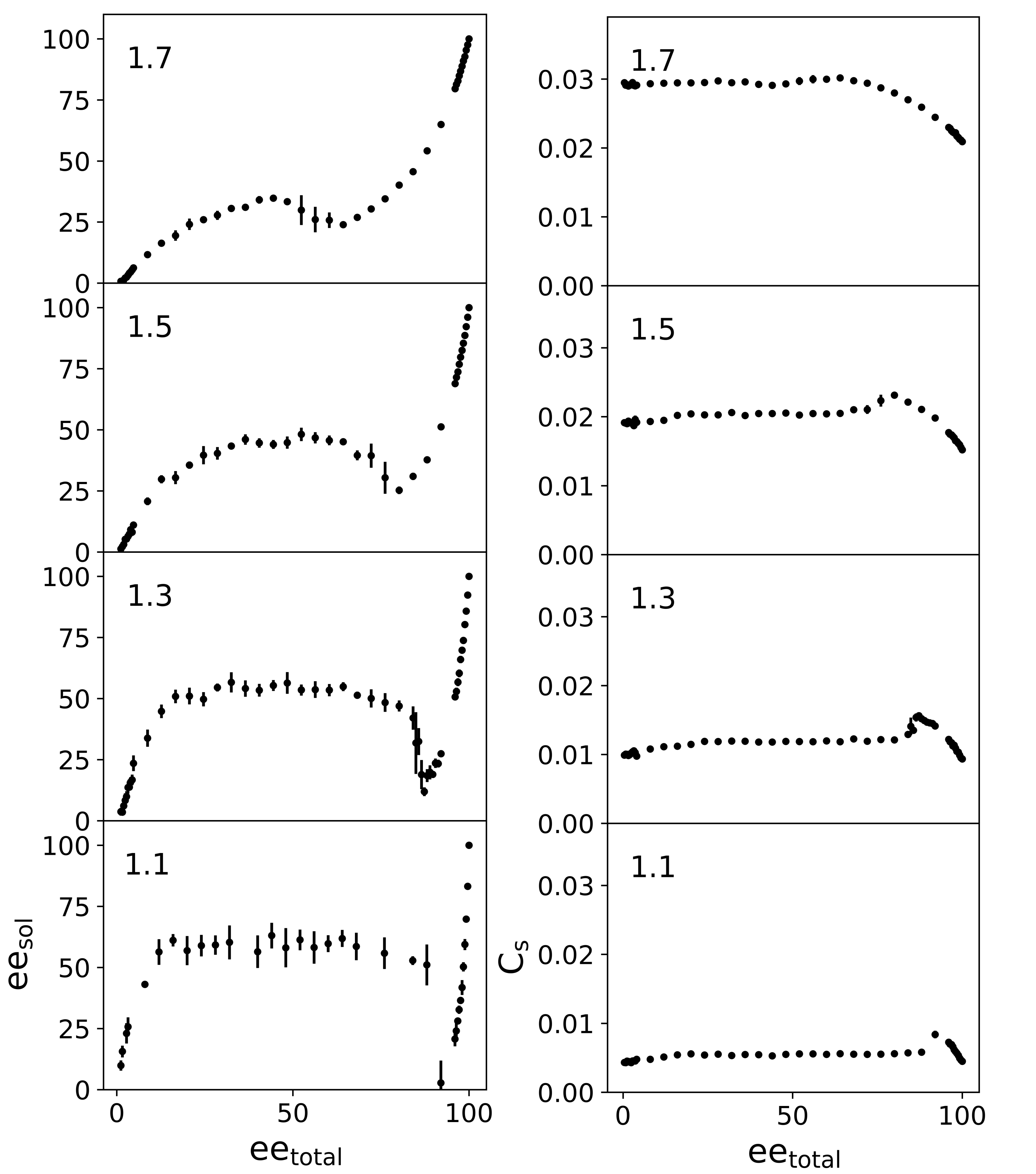}
	\caption{Effect of temperature on enantiomeric excess using the constant $\nu_1 = -2$.}
	\label{Exceso_efect_temperatura}
\end{figure}

It is shown in Figure \ref{Exceso_efect_temperatura} that as the total enantiomeric excess gradually increases, the enantiomeric excess in solution follows a non-linear behavior. There is a more pronounced chiral amplification as the temperature decrease. After a short interval of growth of enantiomeric excess in solution there is a region of constant enantiomeric excess. This behavior has been defined by Klussmann as a region where a eutectic point has been reached, understood as a region in which a solution transforms into two differentiated solids. Following the phase rule $F = 1-P+C$ for the system with three components and three phases, we would have $F = 1-3+3 = 1$, so once the temperature is defined, this concentration in solution will be defined. Passing this point, the behavior of the enantiomeric excess in solution is highly dependent on temperature. At lower temperatures there is a large decrease in the enantiomeric excess in solution. This decrease becomes less pronounced when the temperature increases. The variation in the concentration of the system can also be seen in the right subfigures. As we have already pointed out in the previous section, at higher temperatures the model reproduces the increase in concentration in solution. Regarding the dependence of the concentration against the variation of the total enantiomeric excess, we can also observe a non-linear behavior. The curves show a maximum peak whose position shifts towards lower values of enantiomeric excess as the temperature increases. This behavior is qualitatively very close to the profiles reported by Klussmann \cite{klussmann2006thermodynamic}.
\medskip

\begin{figure}
	\centering
	\includegraphics[scale=0.55]{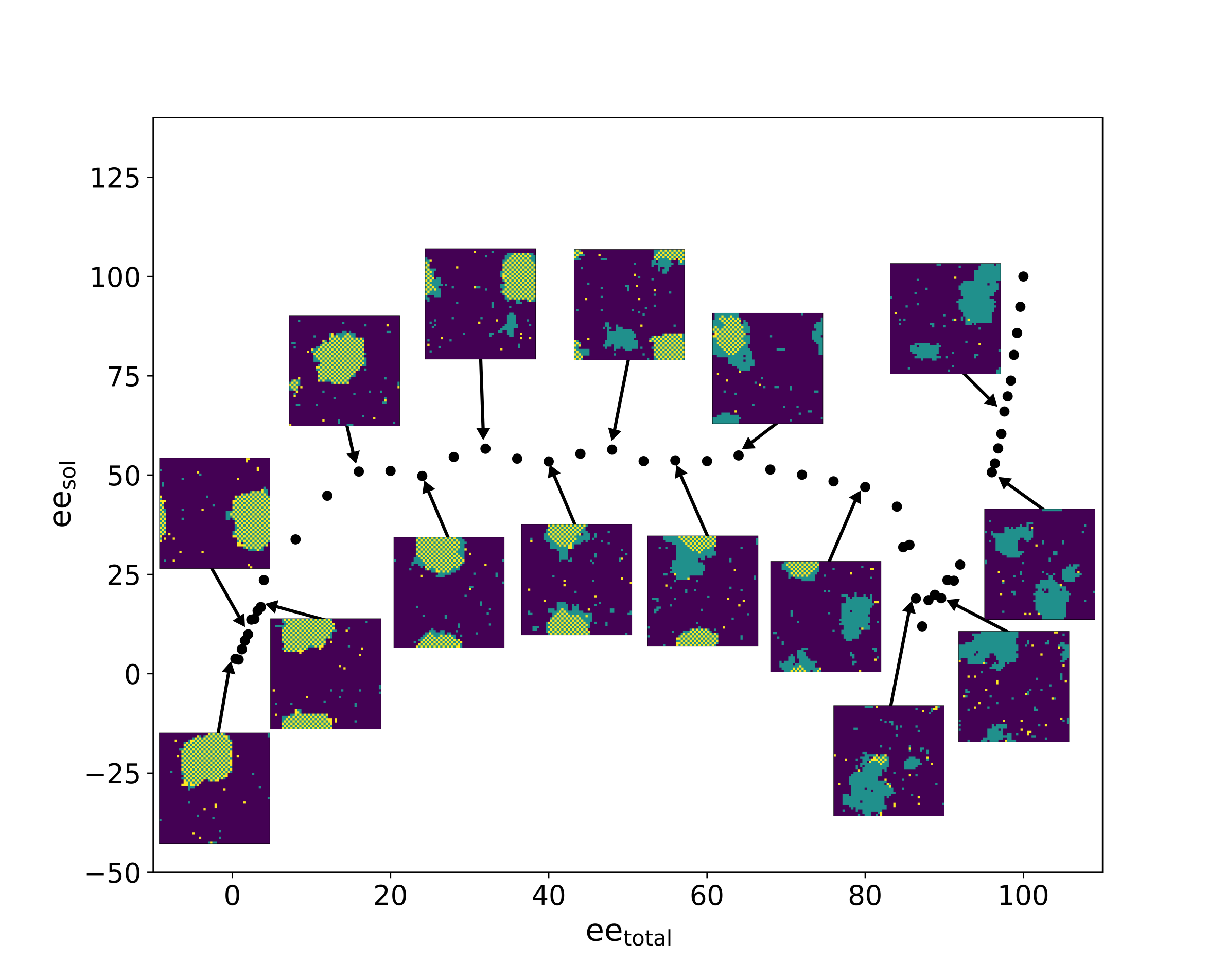}
	\caption{Square lattices showing the crystallization of amino acids in the system at $T = 1.3$ as the total enantiomeric excess increases. The value of the constant $\nu_1$ is $-2$.}
	\label{Exceso_ee_1.3}
\end{figure}

Figure \ref{Exceso_ee_1.3} shows the evolution of the system against the increase of the total enantiomeric excess. We can appreciate that indeed the region of constant enantiomeric excess corresponds to a region where three phases coexist, the solution, the racemic crystal and the enantiopure crystal. The region of steep decline that occurs when the enantiomeric excess is high corresponds to a region where the racemic phase is no longer stable enough to coexist with the enantiopure phase. This causes the enantiomeric excess in solution to decrease. This instability of the racemic crystal could also be associated with an increase in the concentration in solution. When the enantiomeric excess is already very high and close to 100$\%$, the system shows only the enantiopure phase and a dependence of the enantiomeric excess in solution on the $ee_{total}$. In the Supplementary Material section we can see the other lattice for the other temperatures ( Figures S13 to S16)

\subsection{Effect of $\nu_1$ on enantiomeric excess in solution, $ee_{sol}$}

\begin{figure}
	\centering
	\includegraphics[scale=0.52]{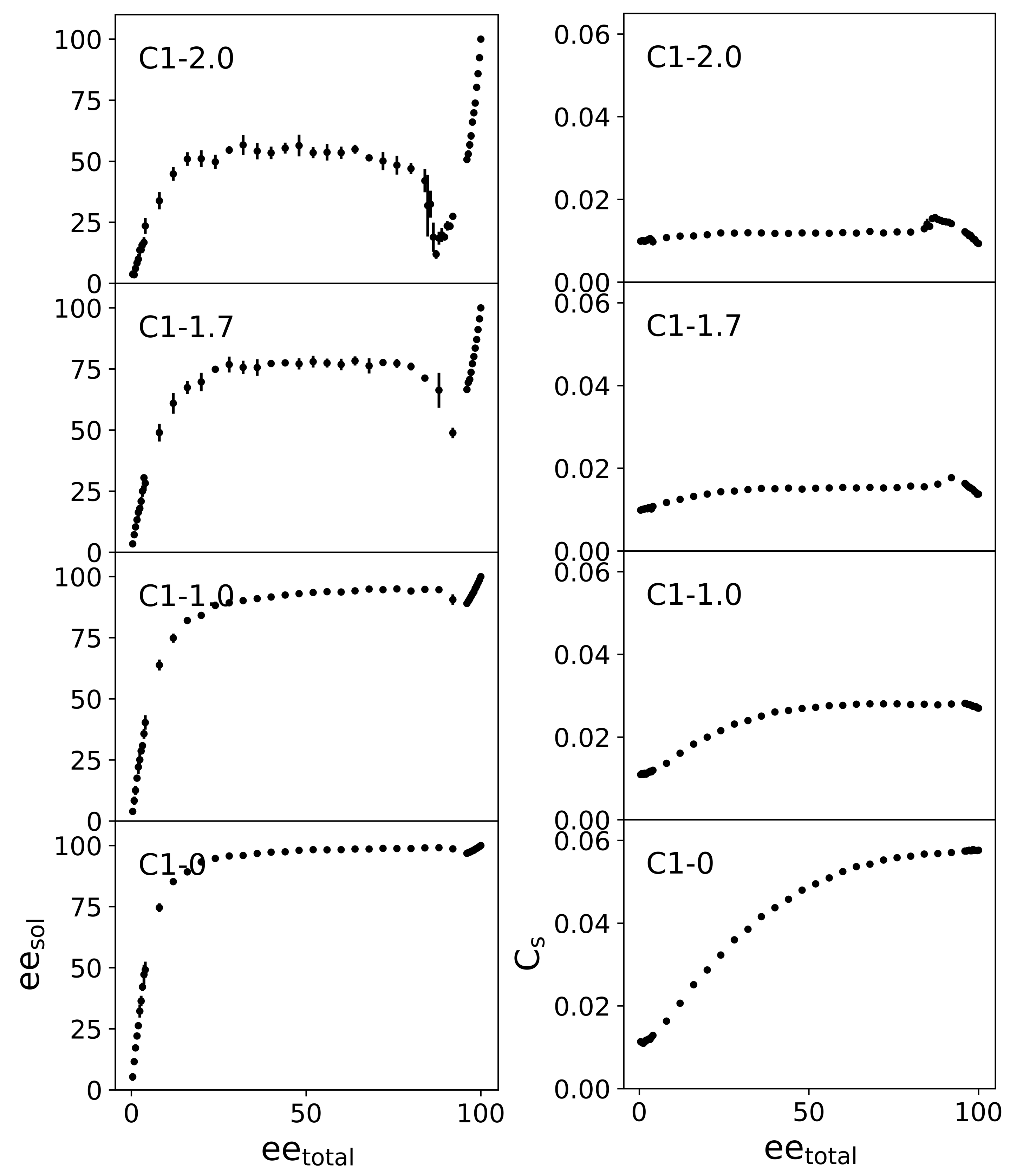}
	\caption{Effect of the value of the constant $\nu_1$ on the enantiomeric excess at the temperature of T = 1.3.}
	\label{Exceso_efect_C1}
\end{figure}

\begin{figure}
	\centering
	\includegraphics[scale=0.55]{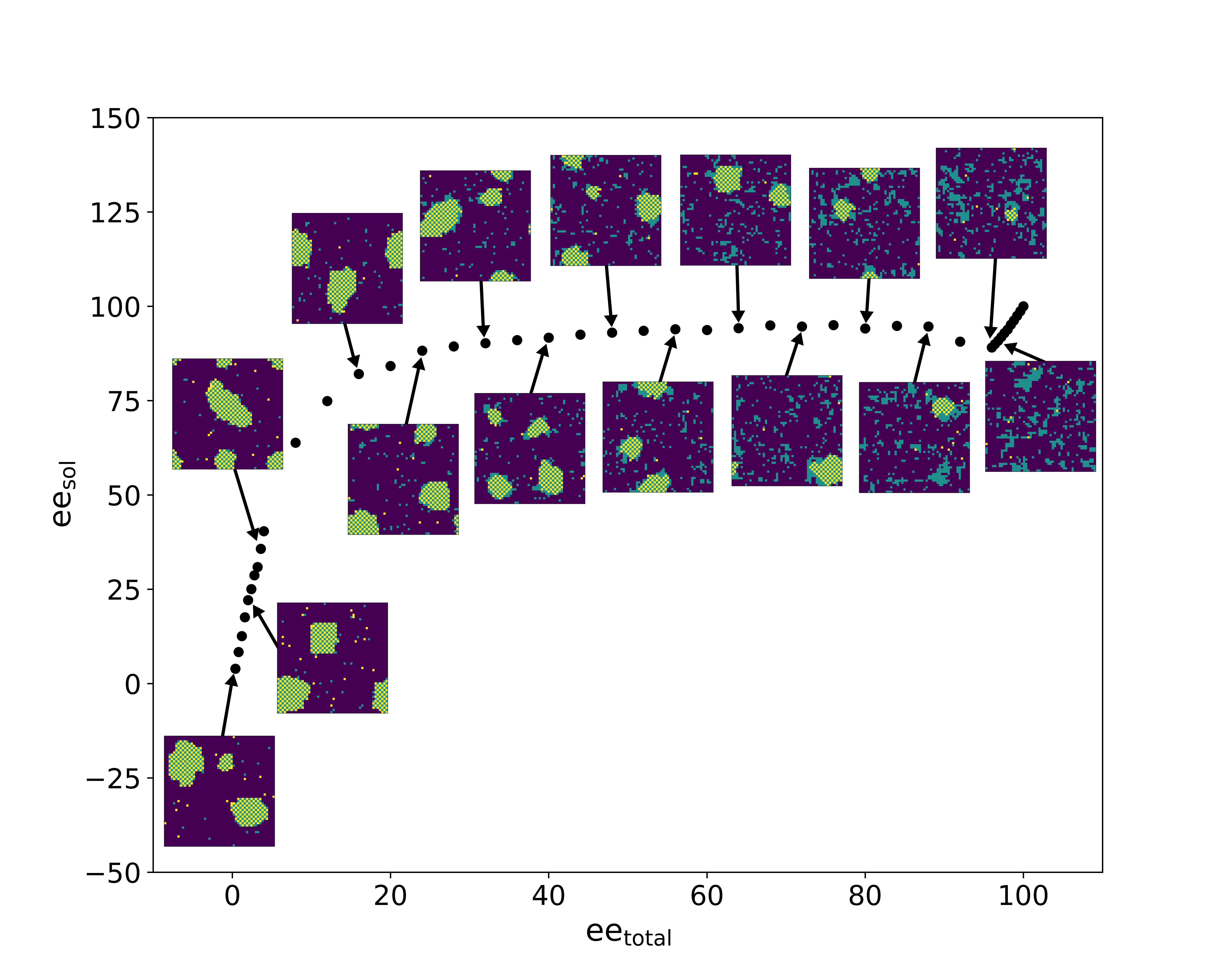}
	\caption{Square lattice showing the crystallization of amino acids in the system at  $T = 1.3$ as the total enantiomeric excess increases. The value of the constant $\nu_1=-1$.}
	\label{Exceso-ee-C1-1.0}
\end{figure}

This effect of enantiomeric excess in solution is also dependent on the value of the constant $\nu_1$ which refers to the energy of formation of an homochiral pair of amino acids. As shown in Figure \ref{Exceso_efect_C1}, as the value of the constant $\nu_1$ decreases the chiral amplification is much more pronounced and the region of declining enantiomeric excess in solution in the region of high total enantiomeric excess is already it's not that drastic. This behavior may be the result of the fact that when this constant decreases, the constant $\nu_2$ plays a more important role in the evolution of the system. Thus, the solid phase evolves towards a racemic state in a very determined way, leaving the enriched solution of the amino acid with more abundant chirality. The concentration in solution also shows a peculiar behavior as we decrease the constant $\nu_1$. This decrease in the value of the constant $\nu_1$ causes the concentration maximum to become less and less noticeable. Figure \ref{Exceso-ee-C1-1.0} shows the evolution of the system when the constant $\nu_1$ is equal to -1. It is shown, as mentioned above, that the passage from a three-phase system (racemic crystal, enantiopure crystal and solution) to a two-phase system is more uniform than when the constant $\nu_1=-2$.  In the Supplementary Material section we can see the other lattice for the other $\nu$ values ( Figures S17 to S20).

\section{Conclusion}
The chiral amplification process that arises by mixing two amino acid enantiomers in a supersaturated system has been studied. The square lattice used, together with the Monte Carlo dynamics, have provided good simulation results and have allowed us to understand the key processes that occur during the formation of amino acid crystals, the variables that determine these processes and the effects of its variation.
\medskip

The square lattice model manages to qualitatively and quantitatively reproduce the phenomenon of supersaturation with increasing concentration in a racemic mixture. It has been found that the solubility fits a van't Hoff solubility curve better than a Hildebrand one. For other hand, the increase in the concentration of amino acids in a racemic mixture produces established clusters that we have identified with established pre-nucleation clusters. We thus conclude that the crystallization process that our model reproduces is a non-classical crystallization process.
\medskip

The simulations that sought to determine the effect of varying the total enantiomeric excess on the enantiomeric excess in solution have made it possible to qualitatively and quantitatively reproduce Klussmann's experimental results. It has been determined that the chiral amplification process owes its origin to the formation of an enantiopure surface phase covering a racemic interior phase together with the preference of the system for the formation of a racemic crystall. Low temperatures favor chiral amplification because an increase in temperature causes a distortion of the racemic phase. The constant that defines the interaction between molecules of the same chirality, $\nu_1$, also plays a preponderant role. By decreasing the value of this constant, the constant $\nu_2$, which defines the stability of the racemic clusters, has a very large influence on the formation of crystals. When $\nu_1$ decreases, the crystals formed are mostly racemic, this causes any enantiomeric excess of the total system to go into solution and this causes an enrichment in one of the enantiomers in solution. On the contrary, by increasing $\nu_1$ with respect to $\nu_2$ it is possible to form distorted racemic crystals or enantipure crystals that allow a balance between the enantiomers present, which causes a lower enantimeric excess in solution.

\section*{Acknowledgements}
RC wishes to acknowledge FONDECYT (Convenio 208-2015-FONDECYT) for his Master scholarship. He would also like to thank Leonardo Leon Vela, Gianfranco Ferro y Miguel Miní for his support in the implementation of the C++ codes.
\bigskip 

{\bf Author contributions} 
Conceptualization: RCS, JCH; Analytical Methodology: RCS, JCH, GP;  Investigation: RCS; Coding and Visualization: RCS; Writing-Original Draft: RCS; Writing-Reviewing and Editing: All authors. 
\bigskip

{\bf Statements and Declarations} The authors declare no conflict of interest.
\medskip


\bigskip

\bibliographystyle{alpha}
\bibliography{pnas-sample}




\end{document}